\documentclass[preprint,5p,times,twocolumn]{elsarticle}
\usepackage{framed} 
\usepackage{multicol} 
\usepackage{nomencl} 
\makenomenclature
\renewcommand*\nompreamble{\begin{multicols}{2}}
\renewcommand*\nompostamble{\end{multicols}}
\usepackage{amsmath}
\usepackage{amssymb}
\usepackage{arydshln} 
\usepackage{amsthm}
\usepackage{graphicx}
\usepackage{subfig}
\usepackage{color}
\usepackage{url}
\definecolor{lightblue}{RGB}{0,176,240}
\definecolor{purple}{RGB}{153,102,255}
\usepackage{amsmath}
\usepackage[scr=boondox,  
            cal=esstix]   
           {mathalpha}
\usepackage{algorithm}
\usepackage{algpseudocode}
\usepackage{float}
\usepackage{multirow}
\usepackage[utf8]{inputenc}
\usepackage[colorlinks,linkcolor=blue, anchorcolor=blue, citecolor=blue]{hyperref}
\usepackage[switch]{lineno}
\journal{ISPRS Journal of Photogrammetry and Remote Sensing}
\begin{document}

\begin{frontmatter}
\title{LSwinSR: UAV Imagery Super-Resolution based on Linear Swin Transformer}

\author[label1]{Rui Li}
\ead{rui.li.4@warwick.ac.uk}
\address[label1]{Intelligent Control \& Smart Energy (ICSE) Research Group, School of Engineering, University of Warwick, Coventry, CV4 7AL, UK}
\author[label1]{Xiaowei Zhao\corref{cor2}}
\ead{xiaowei.zhao@warwick.ac.uk}
\cortext[cor2]{Corresponding author}

\begin{abstract}
Super-resolution, which aims to reconstruct high-resolution images from low-resolution images, has drawn considerable attention and has been intensively studied in computer vision and remote sensing communities. The super-resolution technology is especially beneficial for Unmanned Aerial Vehicles (UAV), as the amount and resolution of images captured by UAV are highly limited by physical constraints such as flight altitude and load capacity. In the wake of the successful application of deep learning methods in the super-resolution task, in recent years, a series of super-resolution algorithms have been developed. In this paper, for the super-resolution of UAV images, a novel network based on the state-of-the-art Swin Transformer is proposed with better efficiency and competitive accuracy. Meanwhile, as one of the essential applications of the UAV is land cover and land use monitoring, simple image quality assessments such as the Peak-Signal-to-Noise Ratio (PSNR) and the Structural Similarity Index Measure (SSIM) are not enough to comprehensively measure the performance of an algorithm. Therefore, we further investigate the effectiveness of super-resolution methods using the accuracy of semantic segmentation. The code will be available at  \url{https://github.com/lironui/GeoSR}.
\end{abstract}

\begin{keyword}
Super-resolution
\sep Transformer
\sep Semantic segmentation
\sep Deep learning
\sep UAV
\end{keyword}
\end{frontmatter}


\section{Introduction}
\label{sec:1}
With the continuous development in both hardware reliability and control strategy, Unmanned Aerial Vehicles (UAV) have been widely employed in more and more practical applications \cite{ZHANG2023109442}, such as atmosphere monitoring \cite{yuan2020maritime}, tracking and surveillance \cite{hu2020joint} and land cover and land use monitoring \cite{xie2021super}. However, limited by the physical characteristics of the UAV such as the flight altitude and load capacity, sometimes it is impractical to continuously obtain High-Resolution (HR) images through UAV, especially for those applications that need large-scale and long-duration UAV video data \cite{mao2022can}. Super-resolution is a promising solution to alleviate this dilemma which can reconstruct high-resolution images from Low-Resolution (LR) observations \cite{dong2015image}. Actually, the super-resolution technology particularly the deep-learning-based methods has already been widely applied for the processing of various kinds of images, such as natural images \cite{lu2022transformer, zhang2021ranksrgan}, medical images \cite{chen2021super, dharejo2022multimodal} and remote sensing images \cite{razzak2023multi, vasilescu2023cnn, wang2022artifact}. These works can not only enhance the image quality but also can further facilitate downstream applications \cite{xiang2022crack, jiang2022super}. 

\begin{table*}[!t]   
\begin{framed}
\nomenclature[a0]{$Abbreviations$}{}
\nomenclature[a1]{Chat-GPT}{Chat Generative Pre-trained Transformer}
\nomenclature[a2]{CNN}{Convolutional Neural Network}
\nomenclature[a3]{EDSR}{Enhanced Deep Super-Resolution Network}
\nomenclature[a4]{FPS}{Frames Per Second}
\nomenclature[a5]{GELU}{Gaussian Error Linear Unit}
\nomenclature[a6]{HR}{High-Resolution}
\nomenclature[b1]{LN}{LayerNorm}
\nomenclature[b2]{LP-KPN}{Laplacian Pyramid based Kernel Prediction Network}
\nomenclature[b3]{LR}{Low-Resolution}
\nomenclature[b4]{LSTL}{Linear Swin Transformer Layers}
\nomenclature[b5]{LSwinSR}{Linear Swin Transformer for Super-Resolution}
\nomenclature[b60]{KA}{Kernel Attention}
\nomenclature[b61]{MACs}{Multiply-Accumulate Operations} 
\nomenclature[b7]{MHKA}{Multi-Head Kernel-Attention}
\nomenclature[b8]{MHSA}{Multi-Head Self-Attention}
\nomenclature[b9]{MLP}{Multi-Layer Perceptron}
\nomenclature[b90]{NLP}{Natural Language Processing}
\nomenclature[b91]{NLSA}{Non-Local Sparse Attention}
\nomenclature[c1]{PSNR}{Peak-Signal-to-Noise Ratio}
\nomenclature[c2]{RLSTB}{Residual Linear Swin Transformer Blocks}
\nomenclature[c3]{SR}{Super-Resolution}
\nomenclature[c4]{SRCNN}{Super-Resolution Convolutional Neural Network}
\nomenclature[c5]{SRGAN}{Super-Resolution Generative Adversarial Network}
\nomenclature[c6]{SRResNet}{Super-Resolution Residual Network}
\nomenclature[c7]{SSIM}{Structural Similarity Index Measure}
\nomenclature[c9]{SwinIR}{Swin Transformer for Image Restoration}
\nomenclature[d1]{UAV}{Unmanned Aerial Vehicles}
\nomenclature[d2]{UNetFormer}{UNet-like Transformer}
\nomenclature[d3]{VDSR}{Very Deep Super Resolution network}
\nomenclature[d4]{ViT}{Vision Transformer}
\nomenclature[d5]{W-MHSA}{Window-based Multi-Head Self-Attention module}
\nomenclature[d9]{}{}
\nomenclature[f0]{$Symbols$}{}
\nomenclature[f01]{$ \boldsymbol B $}{The relative position bias term}
\nomenclature[f1]{$ \mathscr{C}_{DF} $}{The convolutional layer in the deep feature extraction module}
\nomenclature[f2]{$ \mathscr{C}_{RLSTB} $}{The convolutional layer in the residual linear Swin Transformer block}
\nomenclature[f3]{$ \mathscr{C}_{SF} $}{The convolutional layer in the shallow feature extraction module}
\nomenclature[f4]{$ \mathscr{D} $}{The degradation mapping function}
\nomenclature[f5]{$ E $}{The expected loss}
\nomenclature[f6]{$ \mathscr{F} $}{The super-resolution model}
\nomenclature[f7]{$ F_{DF} $}{The deep features}
\nomenclature[f8]{$ F_{SF} $}{The shallow features}
\nomenclature[f80]{$ M $}{The side length of a local window}
\nomenclature[f81]{$ I $}{The number of image pairs}
\nomenclature[f9]{$ \boldsymbol {I}_{LR} $}{The low-resolution image}
\nomenclature[g1]{$ \boldsymbol {I}_{HR} $}{The high-resolution image}
\nomenclature[g2]{$ \boldsymbol {I}_{SR} $}{The super-resolution image}
\nomenclature[h1]{$ L $}{The loss function}
\nomenclature[h2]{$ \mathscr{L} $}{The linear Swin Transformer layer}
\nomenclature[h3]{$ \boldsymbol {Q}, \boldsymbol {K}, \boldsymbol {V} $}{The query, key and value matrices}
\nomenclature[h30]{$ \mathscr R $}{The residual linear Swin Transformer block}
\nomenclature[h4]{$ \mathscr {Re} $}{The reconstruction module}
\nomenclature[h5]{$ \Phi $}{the regularization term}
\nomenclature[h6]{$ \lambda $}{the tradeoff parameter of $ \Phi(\cdot) $}
\nomenclature[h7]{$\theta$}{the parameters of the super-resolution model}
\nomenclature[h8]{$\delta$}{the parameters of the degradation process}
\nomenclature[h9]{$\rho$}{The normalization function}
\printnomenclature
\end{framed}
\end{table*}

Among those revolutionary deep-learning-based super-resolution algorithms, most of them are constructed by the Convolutional Neural Network (CNN) \cite{dong2015image, wang2018esrgan, li2019feedback}. However, as pointed out by \cite{conde2022swin2sr}, the utilization of CNN for super-resolution has two obvious drawbacks. First, as interactions between images and kernels are content-independent, using the same kernel to restore different image regions may not be the optimal solution. Second, as CNN is initially designed to focus on extracting local patterns, it lacks the capability for capturing long-range and non-local dependencies. By contrast, Transformer \cite{vaswani2017attention}, a promising alternative to CNN, adopts the self-attention mechanism to capture global interactions between contexts, which has shown its great potential in boosting vision-related tasks \cite{dosovitskiyimage, liu2021swin, zhudeformable, liu2022swin, wang2022unetformer}. 

For super-resolution, the interactions between images and self-attention blocks in the Vision Transformer (ViT) are content-adaptive as the attention weights are generated according to the relationship between contexts. Meanwhile, the shifted window mechanism embedded within the Transformer enables long-range dependency modelling. However, the utilization of the self-attention mechanism means significant memory and computational costs, which increases quadratically with the size of the input, i.e. $ O(N^2) $ complexity \cite{li2021multiattention}. Even though the shifted window attention operation \cite{liu2021swin, liu2022swin} can partly alleviate the massive memory and computational requirements by only applying the attention operation on a small local window, the quadratic complexity of the self-attention mechanism itself is still a problematic issue of concern, which is especially true when we need a large local window.

Meanwhile, for most super-resolution studies, the performance of the algorithm is only evaluated by the image quality assessment metrics such as the Peak-Signal-to-Noise Ratio (PSNR) and the Structural Similarity Index Measure (SSIM). The image quality assessment metrics may be enough and appropriate for entertaining applications such as enhancing the resolution of antique digital photos. However, when it comes to task-oriented applications such as UAV-based land cover and land use monitoring, those metrics which only reflect the image quality are far from the practical requirements.

In this paper, based on our previous work on linearizing the complexity of the self-attention mechanism \cite{li2021multiattention, li2021multistage, li2021abcnet}, we propose a novel shifted window attention with linear complexity by employing the kernel attention mechanism \cite{li2021multiattention}, thereby designing the Linear Swin Transformer for UAV Super-Resolution (LSwinSR). The experimental results show that the inference speed of the proposed LSwinSR is faster than the Swin Transformer for Image Restoration (SwinIR) \cite{liang2021swinir} but with competitive accuracy. Meanwhile, we compare and evaluate the usability and dependability of the results generated by different super-resolution methods based on semantic segmentation accuracy, which further demonstrates the effectiveness of the proposed LSwinSR.

The remaining part of this paper is organized as follows: the related works are reviewed in Section \ref{sec:2}. Then, the methodology is described in Section \ref{sec:3}. Thereafter, the dataset, experimental setting and experimental result are reported and analyzed in Section \ref{sec:4}. Finally, the conclusions are drawn in Section \ref{sec:5}.

\section{Related Work}
\label{sec:2}

\subsection{Problem Definition}
\label{sec:2.1}
The super-resolution task aims at reconstructing the high-resolution images from the corresponding low-resolution inputs. Generally, the low-resolution image $ \boldsymbol {I}_{LR} $ can be modelled as the output of the degradation process:

\begin{equation}
\boldsymbol {I}_{LR} = \mathscr{D}(\boldsymbol {I}_{HR}; \delta)
\label{equa:1}
\end{equation}

\noindent where $ \mathscr{D} $ denotes a degradation mapping function, $ \boldsymbol {I}_{HR} $ represents the corresponding high-resolution image and $ \delta $ means the parameters of the degradation process. For real-world applications, the degradation process is normally unknown, while the target of the super-resolution model is to reverse the degradation thereby reconstructing the high-resolution image:

\begin{equation}
\boldsymbol {I}_{SR} = \mathscr{F}(\boldsymbol {I}_{LR}; \theta) \approx \mathscr{D}^{-1}(\boldsymbol {I}_{LR}; \theta)\label{equa:2}
\end{equation}

\noindent where $ \boldsymbol {I}_{SR} $ is the image generated by the super-resolution model $ \mathscr{F} $ and $ \theta $ indicates the parameters of the model $ \mathscr{F} $. Hence, a super-resolution model $ \mathscr{F} $ is trained to narrow the gap between the approximation $ \boldsymbol {I}_{SR} $ and the high-resolution reference $ \boldsymbol {I}_{HR} $ as closely as possible by optimizing the parameters $ \theta $:

\begin{gather}
\theta^* =\mathop{\arg\min}\limits_{\theta} E(\theta) \notag \\ 
 E(\theta) =  \sum\nolimits_{i=1}^I L (\boldsymbol {I}_{HR}, \boldsymbol {I}_{SR}) + \lambda \Phi(\theta) \label{equa:3} \\ 
  L (\boldsymbol {I}_{HR}, \boldsymbol {I}_{SR}) =   L (\boldsymbol {I}_{HR}, \mathscr{F}(\boldsymbol {I}_{LR}; \theta)) \notag 
\end{gather}

\noindent where $ E(\theta) $ means the expected loss, the loss function $ L (\boldsymbol {I}_{HR}, \boldsymbol {I}_{SR}) $ measures the disparity between the high-resolution references and the predicted results, $ \Phi(\theta) $ is the regularization term weighted by the tradeoff parameter $ \lambda $, and $ I $ represents the number of image pairs.

\subsection{Super-Resolution}
\label{sec:2.2}
In recent years, benefitting from the significant advances of deep learning, a series of revolutionary methods have been proposed and verified for the super-resolution task. As a pioneering work, a Super-Resolution Convolutional Neural Network (SRCNN) was proposed by Dong et al. \cite{dong2015image}, demonstrating that the traditional sparse-coding-based super-resolution methods can be reformulated into a deep convolutional neural network. In the proposed pipeline, the low-resolution images were upsampled to the same size as the high-resolution references using the Bicubic interpolation whereafter the upsampled images were taken as the input of the network. With only three convolutional layers, the performance of SRCNN outperformed the Bicubic interpolation method by a large margin. However, the over-simplified structure of SRCNN severely limited its full potential for more complicated scenarios. To address this problem, a Very Deep Super Resolution (VDSR) \cite{kim2016accurate} structure was developed with 20 convolutional layers to learn the deep features of the images. Meanwhile, Dong et al. \cite{dong2016accelerating} also further optimized and accelerated their SRCNN by deepening the layer of the network, taking the original low-resolution image as input and adding the deconvolution layer at the end to enlarge the feature map. As the deconvolution operations were prone to checkerboard artefacts, EnhanceNet \cite{sajjadi2017enhancenet} alleviated this problem by replacing the deconvolution layer with the subpixel convolution layer. Besides, the residual block \cite{he2016deep} was also introduced to design super-resolution networks with deeper or wider structures. For example, a Super-Resolution Generative Adversarial Network (SRGAN) was proposed in \cite{ledig2017photo} based on the designed 16 blocks deep Super-resolution ResNet (SRResNet). The SRResNet was further improved by the Enhanced Deep Super-Resolution (EDSR) network \cite{lim2017enhanced}, which removed the batch normalization layers as they would get rid of the range flexibility from the network. Moreover, several studies attempted to combine the advantages of both deep learning and traditional image processing technologies. For example, the Laplacian Pyramid based Kernel Prediction Network (LP-KPN) \cite{cai2019toward}  applied the Laplacian pyramid technology to enlarge the receptive field and reduce the computational cost, while the Non-Local Sparse Attention (NLSA) proposed by \cite{mei2021image} embraced the benefits of sparse representation and non-local operation.

\begin{figure*}[!h]
\centering
\includegraphics[width=18cm]{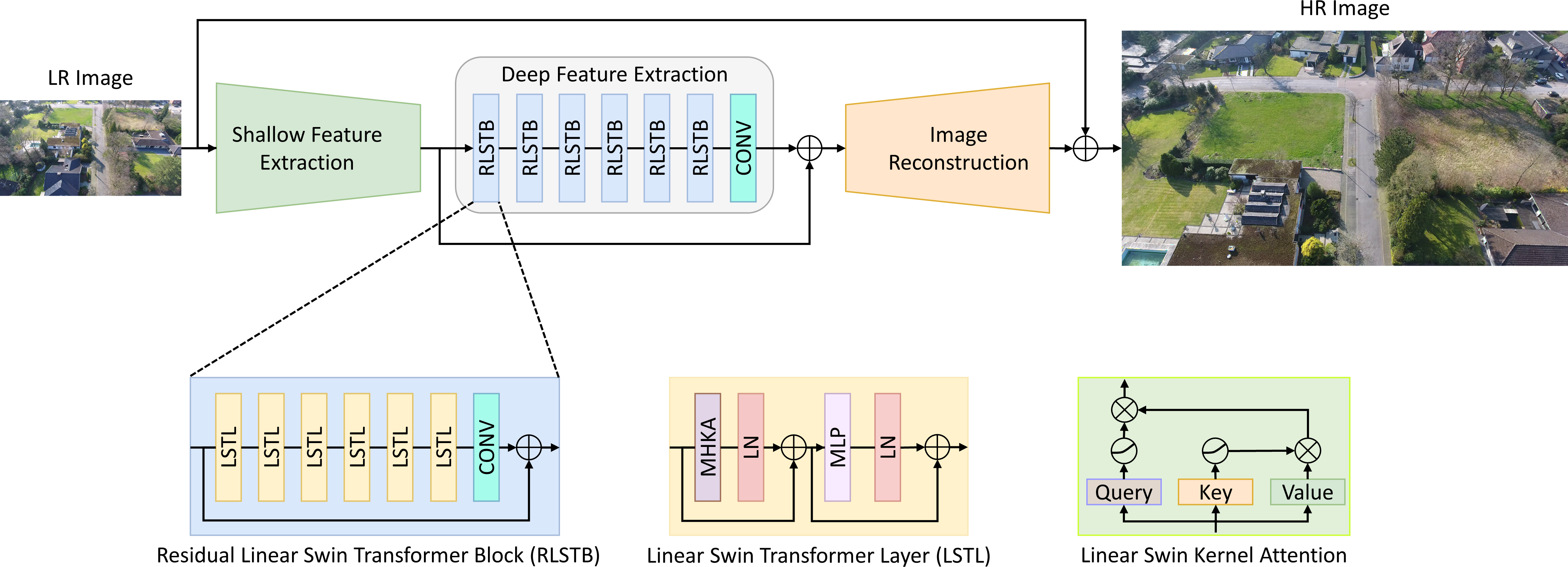}
\caption{The architecture of the proposed LSwinSR for UAV image super-resolution.}
\label{fig:1}
\end{figure*}

\subsection{Vision Transformer}
\label{sec:2.3}
Originally, the Transformer \cite{vaswani2017attention} was designed for Natural Language Processing (NLP) applications, which has achieved state-of-the-art performance and become the de-facto standard solution for many NLP tasks \cite{dosovitskiyimage}, such as the cutting-edge and high-profile Chat Generative Pre-trained Transformer (ChatGPT) \cite{OpenAI2023GPT4TR}. Inspired by the great success of the Transformer in NLP, the Vision Transformer (ViT), a variant of the Transformer designed specifically for image processing, has recently gained much popularity in the computer vision community \cite{conde2022swin2sr, wang2022building}. Different from the CNN structure, the ViT converts 2-D images into 1-D sequences first and then applies the self-attention mechanism for feature extraction. With strong capabilities to capture long-range dependencies and non-local relationships,
the self-attention mechanism can capture complex global interactions between different areas of the image \cite{li2021multiattention}. 

In the state-of-the-art ViT backbone, i.e. Swin Transformer \cite{liu2021swin, liu2022swin}, the Window-based Multi-Head Self-Attention module (W-MHSA) splits the input features into nonoverlapping windows before performing the standard Multi-Head Self-Attention (MHSA) in each local window. Specifically, giving vector $ \boldsymbol X $ as the features within a local window, the self-attention mechanism for each head can be defined as:

\begin{gather}
Attention(\boldsymbol{Q}, \boldsymbol{K}, \boldsymbol{V})=SoftMax(cos(\boldsymbol{Q}, \boldsymbol{K}^T)/\tau + \boldsymbol{B})\boldsymbol{V}\label{equa:4}\\
\boldsymbol {Q = XW_q} \in \mathbb{R}^{M^2 \times d}\notag\\
\boldsymbol {K = XW_k} \in \mathbb{R}^{M^2 \times d}\label{equa:5}\\
\boldsymbol {V = XW_v} \in \mathbb{R}^{M^2 \times d}\notag
\end{gather}

\noindent To calculate the attention map, the \emph{query} matrix $ \boldsymbol Q $, \emph{key} matrix $ \boldsymbol K $ and \emph{value} matrix $ \boldsymbol V $ need to be generated first by applying three projected matrices $ \boldsymbol{W}_q $, $ \boldsymbol{W}_k $ and $ \boldsymbol{W}_v $ to the vector $ \boldsymbol X $, respectively. The generated matrice are all in the shape of $ M^2 \times d $, where $ M^2 $ means the number of pixels in a window and $ d $  indicates the \textit{query/key} dimension. Thereafter, the SoftMax function is applied to each row of the similarity matrix $ cos(\boldsymbol{Q}, \boldsymbol{K}^T)/\tau + \boldsymbol{B} $. Here, the similarity matrix is in the shape of $ M^2 \times M^2 $, which models the relationship between each pair of pixels within the local window. For point $ (i, j) $ in the similarity matrix, the value is computed by:

\begin{equation}
cos(\boldsymbol {q}_i, \boldsymbol {k}_j) + \boldsymbol {B}_{i, j}=(\frac{\boldsymbol {q}_i^T}{\parallel \boldsymbol {q}_i^T \parallel}_2) \cdot (\frac{\boldsymbol {k}_j}{\parallel \boldsymbol {k}_j \parallel}_2) + \boldsymbol {B}_{i, j}\label{equa:6}
\end{equation}

\noindent where $ \boldsymbol {B}_{i, j} $ is the relative position bias between pixel $ i $ and $ j $. $ \tau $ is a learnable scalar, non-shared across heads and layers. Finally, the obtained similarity matrix is multiplied by the \emph{value} matrix $ \boldsymbol V $ to yield the attention map. Compared with the conventional multi-head self-attention module which calculates the similarity matrix among the whole input image, the complexity is significantly reduced as the self-attention operation in Swin Transformer is only conducted within the local window range.

Benefitting from this design, the ViT-based algorithms have demonstrated obvious superiority over CNNs and obtained numerous breakthroughs in fundamental vision tasks, such as image classification \cite{dosovitskiyimage, liu2021swin, liu2022swin}, semantic segmentation \cite{wang2022unetformer, strudel2021segmenter, wang2022novel} and super-resolution \cite{conde2022swin2sr, yang2020learning, liang2021swinir, lei2021transformer}. 

\section{Methodology}
\label{sec:3}
Although the complexity has been already decreased by the window-based multi-head self-attention module to a great extent, the computational and memory requirements for each self-attention operation are still quadratically related to the size of the window. Considering the large number of self-attention operations contained in the Swin Transformer, the optimization for the complexity of the self-attention operation still has considerable benefit in accelerating the inference speed and reducing the memory requirement, especially when we need a large window size to process images in large size.

\subsection{Kernel Attention}
\label{sec:3.1}
Generally, the self-attention mechanism can be formalized as:

\begin{equation}
\begin{aligned}
Attention(\boldsymbol{Q}, \boldsymbol{K}, \boldsymbol{V})&=SoftMax(\boldsymbol{Q} \boldsymbol{K}^T)\boldsymbol{V}\label{equa:7}\\
&=\rho (\boldsymbol{Q} \boldsymbol{K}^T)\boldsymbol{V}
\end{aligned}
\end{equation}

\noindent where $ \rho $ represents the normalization function used to normalize the similarity matrix generated by the product of $ \boldsymbol{Q} $ and $ \boldsymbol{K}^T $. If $ \rho(\cdot) = SoftMax(\cdot) $, the $ i $-th row of the attention map can be written as:

\begin{equation}
Attention(\boldsymbol{Q}, \boldsymbol{K}, \boldsymbol{V})_i = \frac{\sum_{j=1}^M e^{\boldsymbol {q}_i^T \cdot \boldsymbol {k}_j} \boldsymbol{v}_j}{\sum_{j=1}^M e^{\boldsymbol{q}_i^T \cdot \boldsymbol{k}_j}} \label{equa:8}
\end{equation}

As can be drawn from Equation (\ref{equa:8}), the essence of the self-attention mechanism is to weigh the $ \boldsymbol{v}_j $ by $ e^{\boldsymbol {q}_i^T \cdot \boldsymbol {k}_j} $, where $ sim(\boldsymbol {q}_i, \boldsymbol {k}_j)=e^{\boldsymbol {q}_i^T \cdot \boldsymbol {k}_j} $ measures the similarity between the $ \boldsymbol {q}_i $ and $ \boldsymbol {k}_j $. Therefore, Equation (\ref{equa:8}) can be rewritten as:

\begin{equation}\label{equa:9}
\begin{split}
Attention(\boldsymbol{Q}, \boldsymbol{K}, \boldsymbol{V})_i = \frac{\sum_{j=1}^M {\rm sim}({\boldsymbol {q}_i, \boldsymbol {k}_j}) \boldsymbol{v}_j}{\sum_{j=1}^M {\rm sim} ({\boldsymbol{q}_i, \boldsymbol{k}_j})}, where\\
{\rm sim}({\boldsymbol {q}_i, \boldsymbol {k}_j}) \ge 0 \qquad\qquad\qquad
\end{split}
\end{equation}

\noindent Here, $ {\rm sim}({\boldsymbol {q}_i, \boldsymbol {k}_j}) $ can be expanded as $ {\rm sim}({\boldsymbol {q}_i, \boldsymbol {k}_j}) = \phi (\boldsymbol {q}_i)^T \varphi (\boldsymbol {k}_j) $, thereby reforming Equation (\ref{equa:8}) as:

\begin{equation}
Attention(\boldsymbol{Q}, \boldsymbol{K}, \boldsymbol{V})_i = \frac{\sum_{j=1}^M \phi({\boldsymbol {q}_i)^T \varphi(\boldsymbol {k}_j}) \boldsymbol{v}_j}{\sum_{j=1}^M \phi({\boldsymbol {q}_i)^T \varphi(\boldsymbol {k}_j})} \label{equa:10}
\end{equation}

\begin{equation}
Attention(\boldsymbol{Q}, \boldsymbol{K}, \boldsymbol{V})_i = \frac{\phi({\boldsymbol {q}_i)^T\sum_{j=1}^M  \varphi(\boldsymbol {k}_j}) \boldsymbol{v}_j^T}{\phi({\boldsymbol {q}_i)^T\sum_{j=1}^M  \varphi(\boldsymbol {k}_j})} \label{equa:11}
\end{equation}

\noindent In particular, Equation (\ref{equa:10}) is identical to Equation (\ref{equa:8}), when $ {\rm sim}({\boldsymbol {q}_i, \boldsymbol {k}_j}) = e^{\boldsymbol{q}_i^T \cdot \boldsymbol{k}_j} $. Further, Equation (\ref{equa:11}) can be represented as the vectorized form:

\begin{equation}
Attention(\boldsymbol{Q}, \boldsymbol{K}, \boldsymbol{V}) = \frac{{\rm \phi}(\boldsymbol Q){\rm \varphi} (\boldsymbol K)^T \boldsymbol{V}}{{\rm \phi}(\boldsymbol {Q}) \sum_j{{\rm \varphi} (\boldsymbol {K})_{i, j}^T }} \label{equa:12}
\end{equation}

\noindent As the SoftMax function in Equation (\ref{equa:12}) is replaced by $ {\rm sim}({\boldsymbol {q}_i, \boldsymbol {k}_j}) = \phi (\boldsymbol {q}_i)^T \varphi (\boldsymbol {k}_j) $, the order of the operation can be altered. Specifically, we can compute the multiplication between $ {\rm \varphi} (\boldsymbol K)^T $ and $ \boldsymbol{V} $ first and then multiply the result and $ {\rm \phi}  ({\boldsymbol Q}) $, thereby avoiding the intensive computation and generation procedure of the matrix $ \boldsymbol{Q} \boldsymbol{K}^T \in \mathbb{R}^ {M^2 \times M^2} $. Especially, $ \phi(\cdot) $ and $ \varphi(\cdot) $ in Equation (\ref{equa:11}) can be considered as kernel smoothers \cite{tsai2019transformer}. In our previous work \cite{li2021multiattention}, we have shown that by selecting $ {\rm sim}({\boldsymbol {q}_i \boldsymbol {k}_j}) = {\rm SoftPlus}(\boldsymbol {q}_i)^T {\rm SoftPlus}(\boldsymbol {k}_j) $ where $ {\rm SoftPlus}(x) = \log (1+e^x) $, a revised self-attention mechanism (i.e. kernel attention \cite{li2021multiattention}) with linear complexity can be achieved with competitive accuracy. For kernel attention, the Equation (\ref{equa:11}) and Equation (\ref{equa:12}) can be written as:

\begin{equation}
Attention(\boldsymbol{Q}, \boldsymbol{K}, \boldsymbol{V})_i = \frac{{\rm SoftPlus}(\boldsymbol {q}_i)^T \sum_{j=1}^M{{\rm SoftPlus} (\boldsymbol {k}_j) }\boldsymbol{v}_j^T}{{\rm SoftPlus}(\boldsymbol {q}_i)^T \sum_{j=1}^M{{\rm SoftPlus} (\boldsymbol {k}_j) }} \label{equa:13}
\end{equation}

\begin{equation}
Attention(\boldsymbol{Q}, \boldsymbol{K}, \boldsymbol{V}) = \frac{{\rm SoftPlus}(\boldsymbol Q){\rm SoftPlus} (\boldsymbol K)^T \boldsymbol{V}}{{\rm SoftPlus}(\boldsymbol {Q}) \sum_j{{\rm SoftPlus} (\boldsymbol {K})_{i, j}^T }} \label{equa:14}
\end{equation}

\subsection{LSwinSR}
\label{sec:3.2}
The proposed LSwinSR is based on the Swin Transformer \cite{liu2021swin} and SwinIR \cite{liang2021swinir}. As can be seen from Fig. \ref{fig:1}, similar to SwinIR \cite{liang2021swinir}, there are three components included in the LSwinSR, i.e. Shallow Feature Extraction, Deep Feature Extraction and Image Reconstruction.

\begin{figure}[]
\centering
\includegraphics[width=9cm]{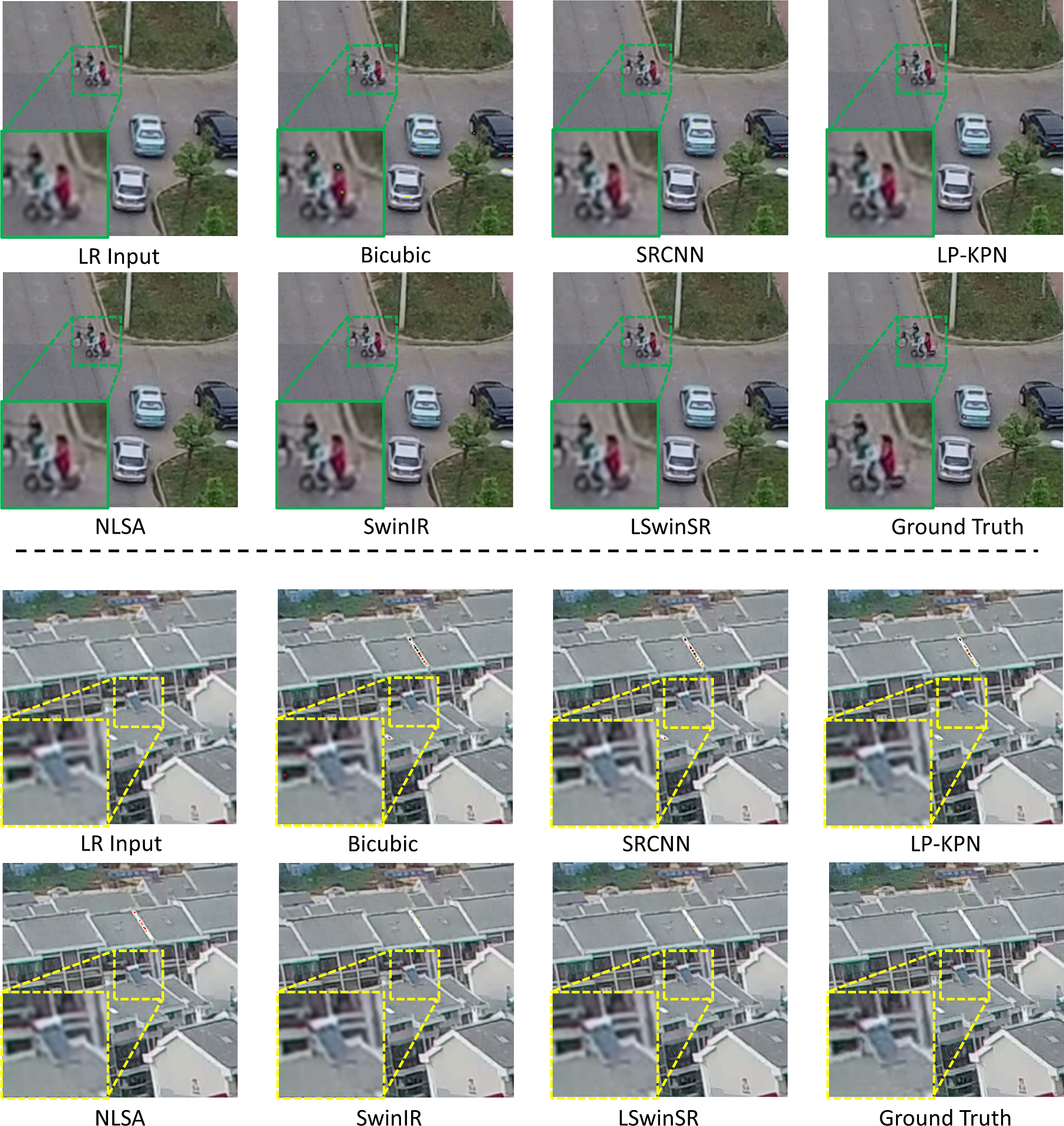}
\caption{Visual comparison of SR ($ \times 2 $) images between different methods on training (top) and validation (bottom) sets for segmentation.}
\label{fig:2}
\end{figure}

\subsubsection{Shallow Feature Extraction}
\label{sec:3.2.1}
Giving a low-resolution input image $ \boldsymbol {I}_{LR} \in \mathbb{R}^ {H \times W \times C} $, a $ 3 \times 3 $ convolutional layer $ \mathscr {C}_{SF} $ is first applied to extract the shallow feature $ \boldsymbol {F}_{SF} \in \mathbb{R}^ {H \times W \times D} $:

\begin{equation}
\boldsymbol {F}_{SF} = \mathscr{C}_{SF}(\boldsymbol {I}_{LR})
\label{equa:15}
\end{equation}

\noindent where $ H $, $ W $, $ C $ and $ D $ are the image height, image width, input channel number and feature channel number, respectively. As pointed out by \cite{xiao2021early}, the utilization of the convolution layer at the early visual processing is beneficial for stable optimization and better results. Besides,  the convolution operation allows mapping the input image space to a higher dimensional feature space in a simple way.

\subsubsection{Deep Feature Extraction}
\label{sec:3.2.2}
After obtaining the shallow feature, the deep features are extracted sequentially by the deep feature extraction module with $ H $ Residual Linear Swin Transformer Blocks (RLSTB) $ \mathscr {R} $ and a $ 3 \times 3 $ convolutional layer $ \mathscr {C}_{DF} $:

\begin{gather}
\boldsymbol {F}_{DF} = \mathscr{C}_{DF}(\boldsymbol {F}_H), where \notag\\
\boldsymbol {F}_i = \mathscr{R}_i(\boldsymbol {F}_{i - 1}), i = 1,2,\ldots,H  \label{equa:16}\\ 
\boldsymbol {F}_0 = \boldsymbol {F}_{SF} \notag
\end{gather}

\noindent The convolutional layer at the end of the deep feature extraction module can provide the inductive bias to the Transformer-based network and pave a better way for the aggregation of shallow and deep features \cite{liang2021swinir}.

\begin{figure*}[]
\centering
\includegraphics[width=18cm]{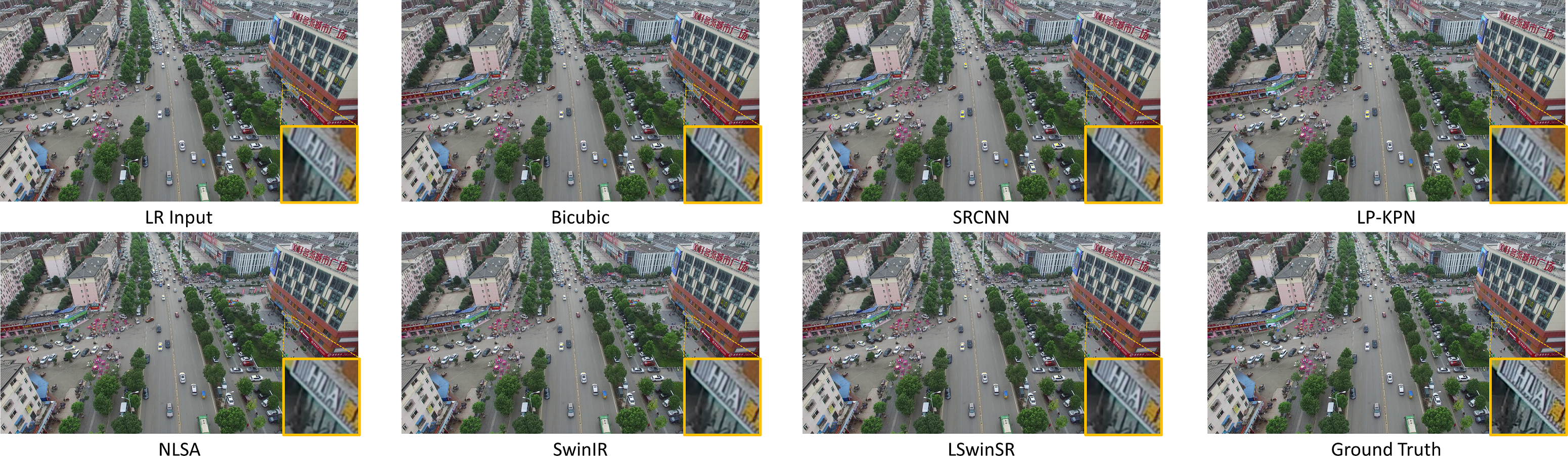}
\caption{Visual comparison of SR ($ \times 2 $) images between different methods on the test set.}
\label{fig:3}
\end{figure*}

As can be seen from Fig. \ref{fig:1}, a residual linear Swin Transformer block comprises $ K $ Linear Swin Transformer Layers (LSTL) $ \mathscr {L} $ and a convolutional layer $ \mathscr {C}_{RLSTB} $. Assuming the input feature maps of the $ i $-th RLSTB is $ \boldsymbol {F}_{i, 0} $, then the output of the current RLSTB $ \boldsymbol {F}_{i, out} $ can be represented as:

\begin{equation}\label{equa:17}
\begin{split}
\boldsymbol {F}_{i, out} = \mathscr{C}_{RLSTB}(\boldsymbol {F}_{i, K}) + (\boldsymbol {F}_{i, 0}), where \\
\boldsymbol {F}_{i, j} = \mathscr{L}_j (\boldsymbol {F}_{i, j - 1}), j = 1,2,\ldots,K \quad
\end{split}
\end{equation}

\noindent Especially, the short identity-based connection provided by the residual connection enables the aggregation of different levels of features.

For each linear Swin Transformer layer, there exist two residual connections and four layers, i.e. a Multi-Head Kernel Attention (MHKA) operation, a Multi-Layer Perceptron (MLP) and two LayerNorm (LN) operations. The LSTL first reshapes the input from the size of $ H \times W \times D $ to the $ HW/M^2 \times M^2 \times D $ by partitioning the input into non-overlapping $ M \times M $ local windows, where $ HW/M^2 $ equals the total number of windows.  Then, the MHKA operation is applied separately for each window, while the technical details have been illustrated in Section \ref{sec:2.3} and Section \ref{sec:3.1}. Thereafter, a MLP with two fully connected layers and a Gaussian Error Linear Unit (GELU) between them is employed for further feature transformations. Meanwhile, the LN layer and the residual connection are adopted for both MHKA and MLP. The whole process of the LSTL can be formulated as:

\begin{gather}
\boldsymbol {X} = MHKA(LN(\boldsymbol {X})) + \boldsymbol {X}\label{equa:18}\\
\boldsymbol {X} = MLP(LN(\boldsymbol {X})) + \boldsymbol {X}\label{equa:19}
\end{gather}

\noindent Meanwhile, to provide connections across local windows, the regular and shifted window partitioning \cite{liu2021swin} is used which shifts the feature by $ (M/2, M/2) $ pixels before partitioning.

\subsubsection{Image Reconstruction}
\label{sec:3.2.3}

To reconstruct the high-resolution images $ \boldsymbol {I}_{SR} $, the shallow features $ \boldsymbol {F}_{SF} $ and deep features $ \boldsymbol {F}_{DF} $ are aggregated as:

\begin{equation}
\boldsymbol {I}_{SR} = \mathscr{Re}(\boldsymbol {F}_{SF} + \boldsymbol {F}_{DF})
\label{equa:20}
\end{equation}

\noindent where $ \mathscr{Re} $ is the reconstruction module. The reason why a long skip connection exists here is that deep features mainly focus on recovering high-frequencies while shallow features primarily centre on reconstructing low-frequencies \cite{liang2021swinir}. With the long skip connection, the low-frequency information contained in shallow features can be directly fed into the reconstruction module, enabling the deep feature extraction module to mainly extract high-frequency information.

\begin{table*}[]
\setlength{\abovecaptionskip}{0.cm}
\centering
\caption{Quantitative comparison including PSNR, SSIM (\%) and MAE (\%) with different methods on the training and validation sets for segmentation and the test set, where the best result is \textbf{highlighted} while the second best is \underline{underlined}.}
\label{table:1}
\begin{tabular}{ccccccccccc}
\hline
\multirow{2}{*}{Models} & \multirow{2}{*}{Scale} & \multicolumn{3}{c}{Train}     & \multicolumn{3}{c}{Validation} & \multicolumn{3}{c}{Test} \\
&  & PSNR $\uparrow$  & SSIM $\uparrow$ & MAE $\downarrow$ & PSNR  $\uparrow$  & SSIM $\uparrow$ & MAE $\downarrow$ & PSNR $\uparrow$  & SSIM $\uparrow$ & MAE $\downarrow$ \\ \hline
Bicubic                 & $ \times 2 $           & 28.760 & 90.878    & 2.273    & 28.817  & 90.863    & 2.329    & 27.933 & 85.956    & 2.437    \\
SRCNN \cite{dong2015image}                  & $ \times 2 $           & 32.994 & 93.195    & 1.986    & 32.907  & 93.079    & 2.045    & 30.516 & 83.791    & 2.859    \\
LP-KPN \cite{cai2019toward}               & $ \times 2 $           & 33.782 & 93.719    & 1.832    & 33.722  & 93.579    & 1.891    & 31.089 & 87.668    & 2.551    \\
NLSA \cite{mei2021image}                & $ \times 2 $           & 34.033 & 93.969    & 1.779    & 33.950  & 93.815    & 1.839    & 31.412 & 88.571    & 2.469    \\
SwinIR \cite{liang2021swinir}                 & $ \times 2 $           & \textbf{34.354} & \textbf{94.240}    & \textbf{1.720}    & \textbf{34.253}  & \textbf{94.073}    & \textbf{1.782}    & \underline{31.997} & \textbf{89.957}    & \underline{2.305}    \\
LSwinSR                 & $ \times 2 $           & \underline{34.277} & \underline{94.169}    & \underline{1.738}    & \underline{34.173}  & \underline{94.004}    & \underline{1.801}    & \textbf{32.040} & \underline{89.868}    & \textbf{2.303}    \\ \hdashline
Bicubic                 & $ \times 4 $           & 22.915 & 66.157    & 4.862    & 22.904  & 66.022    & 4.991    & 23.967 & 69.091    & 3.815    \\
SRCNN \cite{dong2015image}                  & $ \times 4 $           & 25.296 & 67.972    & 4.250    & 25.275  & 67.671    & 4.388    & 24.938 & 54.851    & 8.662    \\
LP-KPN \cite{cai2019toward}               & $ \times 4 $           & 25.960 & 71.908    & 3.888    & 25.887  & 71.431    & 4.037    & 27.963 & 75.774    & \underline{3.668}    \\
NLSA \cite{mei2021image}                   & $ \times 4 $           & 26.117 & 72.182    & 3.914    & 26.013  & 71.628    & 4.069    & 27.524 & 73.330    & 4.212    \\
SwinIR \cite{liang2021swinir}                  & $ \times 4 $           & \textbf{26.338} & \textbf{73.563}    & \textbf{3.731}    & \textbf{26.253}  & \textbf{73.040}    & \textbf{3.872}    & \textbf{28.150} & \textbf{76.505}    & \textbf{3.628}    \\
LSwinSR                 & $ \times 4 $           & \underline{26.227} & \underline{73.075}    & \underline{3.777}    & \underline{26.134}  & \underline{72.558}    & \underline{3.928}    & \underline{27.975} & \underline{76.202}    & 3.687    \\ \hdashline
Bicubic                 & $ \times 8 $           & 19.923 & 46.763    & 6.930    & 19.920  & 46.740    & 7.094    & 21.044 & 50.527    & 5.562    \\
SRCNN \cite{dong2015image}                  & $ \times 8 $           & 21.581 & 48.912    & 6.175    & 21.614  & 48.865    & 6.320    & 20.000 & 34.340    & 23.075   \\
LP-KPN \cite{cai2019toward}               & $ \times 8 $           & 22.151 & 51.735    & 5.653    & 22.159  & 51.538    & 5.819    & 25.787 & \underline{67.759}    & \textbf{5.124}    \\
NLSA \cite{mei2021image}                   & $ \times 8 $           & \textbf{22.886} & \textbf{55.511}    & \textbf{5.223}    & \textbf{22.859}  & \textbf{55.117}    & \textbf{5.385}    & 18.453 & 37.480    & 26.185   \\
SwinIR \cite{liang2021swinir}                 & $ \times 8 $           & 22.402 & 52.770    & 5.501    & 22.434  & 52.645    & 5.642    & \underline{26.144} & \textbf{67.954}    & \underline{5.303}    \\
LSwinSR                 & $ \times 8 $           & \underline{22.494} & \underline{53.493}    & \underline{5.438}    & \underline{22.506}  & \underline{53.265}    & \underline{5.588}    & \textbf{26.256} & 66.269    & 5.683   \\ \hline
\end{tabular}
\end{table*}

\section{Results and discussions}
\label{sec:4}

\subsection{Experimental Setting}
\subsubsection{Dataset}

To compare the performance between different methods, we conduct a series of experiments based on a high-resolution UAV semantic segmentation dataset, i.e. the UAVid dataset \cite{lyu2020uavid}. The UAVid dataset focuses on urban street scenes with 4K resolutions ($ 3840 \times 2160 $ or $ 4096 \times 2160 $) and eight classes. Both the super-resolution and segmentation of the UAVid dataset are extremely challenging because of the high resolution, heterogeneous spatial variation and generally complex scenes. 

UAVid has 42 sequences with a total of 420 images in the dataset (10 images in each sequence), where 27 sequences are used for training and validation and 15 sequences are officially provided without publicly available labels for testing. In our experiments, 15 sequences in the test set are remained for testing the performance of both the super-resolution and the segmentation. For the training and validation set, the images within a sequence are divided into four parts: images with ID: 000000, 000200, 000400, 000600 and the image with ID: 000800 are used for training and validating the super-resolution model while images with ID: 000100, 000300, 000500, 000700 and the image with ID: 000900 are used for training and validating the segmentation model, respectively. In such a setting, the datasets for super-resolution and for segmentation are totally separated, thereby ensuring an independent and objective evaluation.

\subsubsection{Model Training}
\label{sec:4.1.2}

Based on the UAVid dataset, the super-resolution for three kinds of upsampling scales, i.e. $ \times 2 $, $ \times 4 $ and $ \times 8 $, are carried out. For those three scales, the high-resolution images are first cropped into $ 256 \times 256 $, $ 512 \times 512 $ or $ 1024 \times 1024 $ patches, respectively. Then, based on the resize function defined in the OpenCV \cite{opencv_library}, the cropped patches are downsampled to $ 1 / 2 $, $ 1 / 4 $ or $ 1 / 8 $ of the original sizes, i.e. the low-resolution images for three experiments are both in $ 128 \times 128 $. Before the downsampling procedure of the test set, we apply the Gaussian Blur to raw high-resolution images first, in order to simulate the practical usage scenario where the degradation procedure of the test set is normally unknown and different from the experimental setting of the training set. Finally, the super-resolution models are trained to reconstruct the corresponding $ \times 2 $, $ \times 4 $ or $ \times 8 $ high-resolution images. For model training, the $ L_1 $ pixel loss is selected as the loss function:

\begin{equation}
L (\boldsymbol {I}_{HR}, \boldsymbol {I}_{SR}) =  {\parallel \boldsymbol {I}_{HR} - \boldsymbol {I}_{SR} \parallel}_1
\label{equa:21}
\end{equation}

\subsubsection{Model Evaluation}

After training and validation, the super-resolution model is then employed to upsample the training, validation and test sets for segmentation. To evaluate the performance of the super-resolution results, three frequently-used metrics are adopted including the Peak-Signal-to-Noise Ratio (PSNR), the Structural Similarity Index Measure (SSIM) and the Mean Absolute Error (MAE). Thereafter, we train, validate and test a segmentation model, i.e. the UNet-like transformer (UNetFormer) \cite{wang2022unetformer} based on the upsampled images generated from different super-resolution algorithms. By observing, comparing and analysing the segmentation accuracy, a more application-oriented evaluation for super-resolution can be achieved.

\begin{figure}[!h]
\centering
\includegraphics[width=9cm]{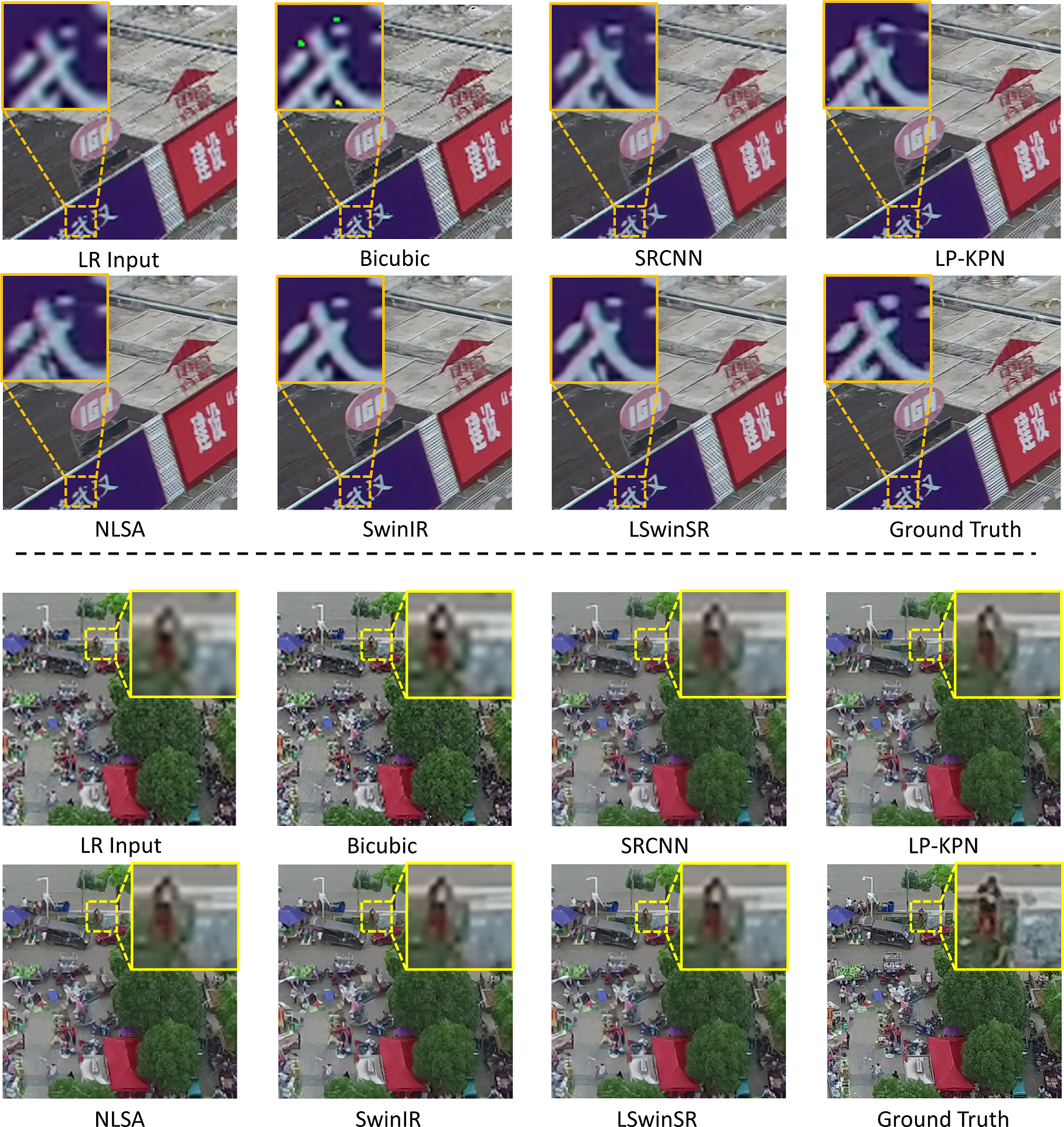}
\caption{Visual comparison of SR ($ \times 4 $) images between different methods on the training (top) and validation (bottom) sets for segmentation.}
\label{fig:4}
\end{figure}

\begin{figure*}[]
\centering
\includegraphics[width=18cm]{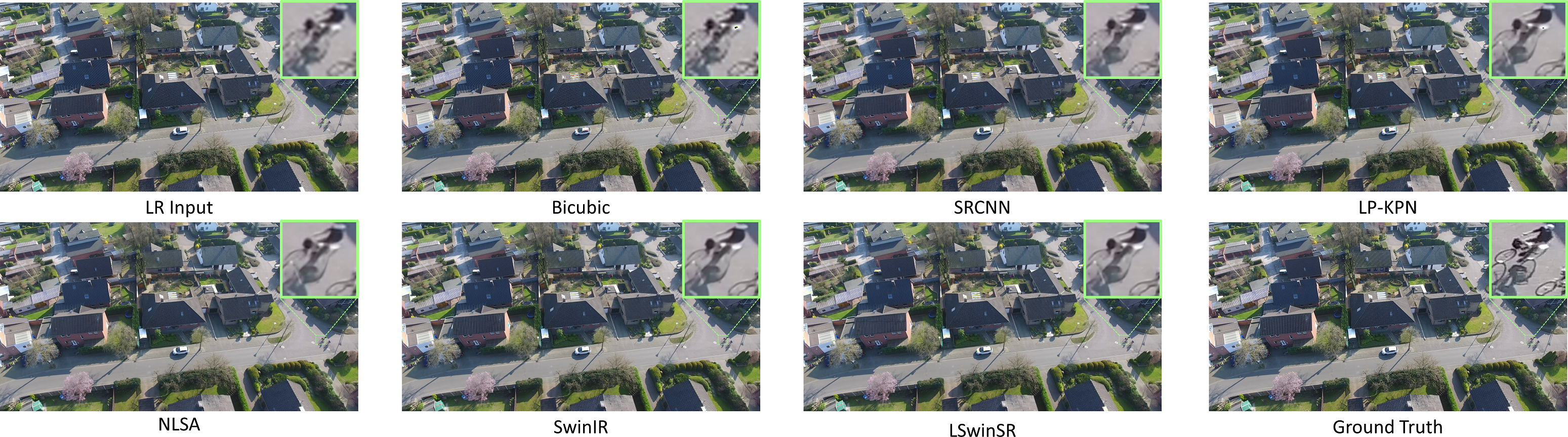}
\caption{Visual comparison of SR ($ \times 4 $) image between different methods on the test set.}
\label{fig:5}
\end{figure*}

\subsection{Super-resolution Performance}
In this section, the performance of six different super-resolution methods including Bicubic, SRCNN \cite{dong2015image}, LP-KPN \cite{cai2019toward}, NLSA \cite{mei2021image}, SwinIR \cite{liang2021swinir} and the proposed LSwinSR is evaluated based on the image quality assessments, i.e. PSNR, SSIM and MAE. Among these six methods, the SRCNN \cite{dong2015image}, LP-KPN \cite{cai2019toward} and NLSA \cite{mei2021image} are CNN-based models, while the LP-KPN \cite{cai2019toward} and NLSA \cite{mei2021image} enhance the model performance by combining the traditional image processing technologies. The SwinIR \cite{liang2021swinir} and the proposed LSwinSR, by contrast, are both constructed based on the state-of-the-art Swin Transformer \cite{liu2021swin}. The SwinIR \cite{liang2021swinir} and the proposed LSwinSR in this section are both based on the lightweight version. Please note that the training and validation sets in this section refer to the training and validation sets for the semantic segmentation which is totally separated from the training and validation sets for super-resolution.

\subsubsection{Quantitative Results} 

As explained in Section \ref{sec:4.1.2}, we add the Gaussian Blur to the test set before generating the low-resolution images. Therefore, the probability distributions between the training set and the test set are different, which means that untrained patterns exist when predicting the test set. This is part of the reason why the performance of SRCNN \cite{dong2015image} is better than the Bicubic interpolation on the training and validation sets but worse on the test set. To be specific, the SRCNN \cite{dong2015image} takes the Bicubic upsampled images as the input (i.e. pre-upsampling) and then processes the input based on only three convolutional layers. Thus, the errors that existed in the Bicubic upsampled images will be further enlarged by the post-processing procedures, which is especially true when considering that there are only three convolutional layers in the SRCNN \cite{dong2015image} to process the upsampled images. By contrast, also as a pre-upsampling-based network, the enhanced post-processing procedures in the LP-KPN \cite{cai2019toward} guarantee a more stable and more robust performance on the untrained test set. 

NLSA \cite{mei2021image}, SwinIR \cite{liang2021swinir} and the proposed LSwinSR are all based on the mainstream post-upsampling structure, which extracts the feature maps in the low-resolution input and upsamples the features at the end. A significant advantage of the post-upsampling structure is the much lower computational cost as the feature extraction procedure is only conducted on the low-resolution space. Thus, the more complex and advanced structure becomes acceptable and practical. For example, the NLSA \cite{mei2021image} holds better performance than the pre-upsampling-based network most of the time. However, for the $ \times 8 $ scale, a severe over-fitting problem occurs in the NLSA \cite{mei2021image}. The accuracy of the NLSA \cite{mei2021image} for $ \times 8 $ scale is the worst among all six methods, although with the best performance on both train and validation sets.

\begin{figure}[htb]
\centering
\includegraphics[width=9cm]{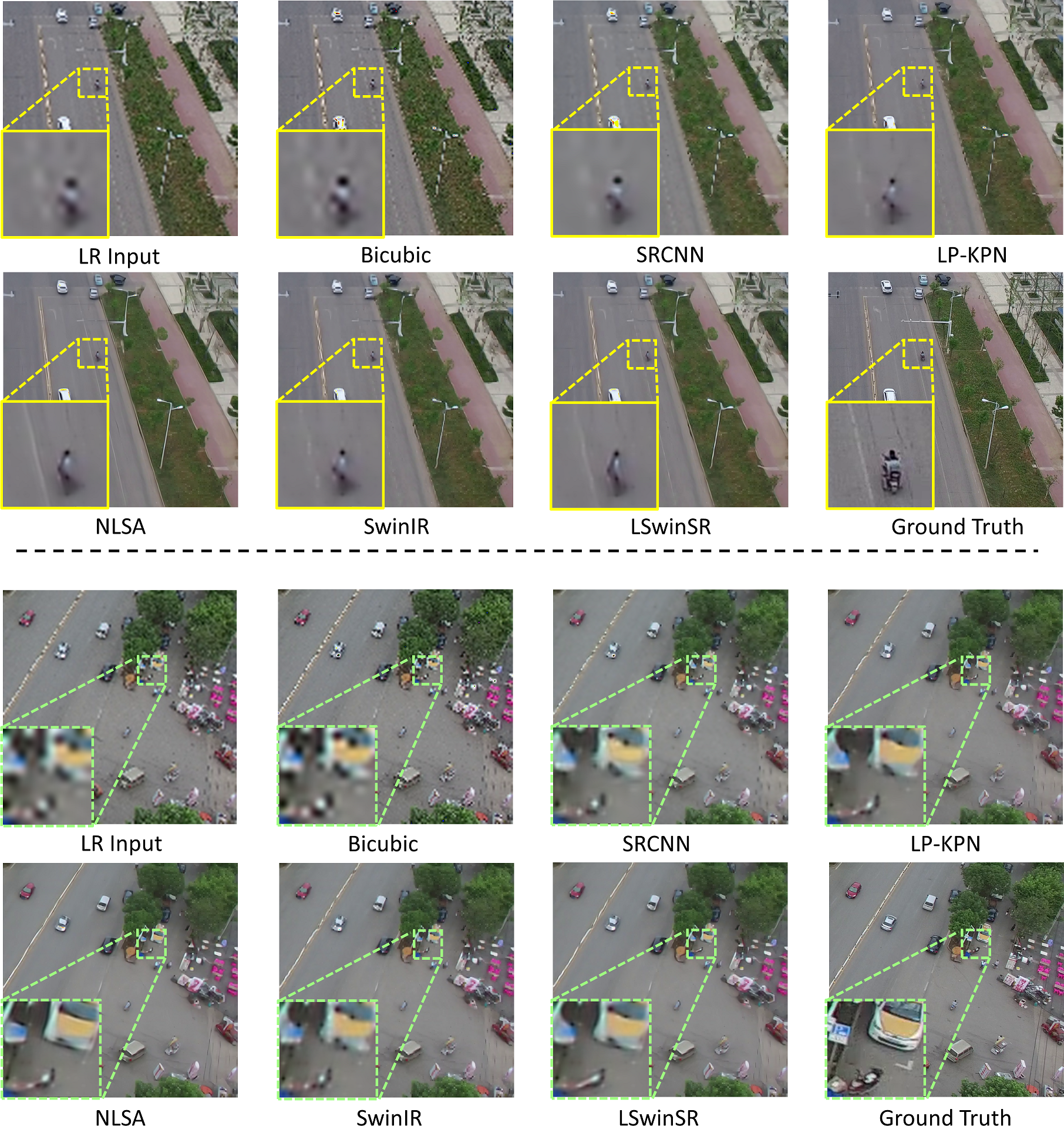}
\caption{Visual comparison of SR ($ \times 8 $) images between different methods on the training (top) and validation (bottom) sets for segmentation.}
\label{fig:6}
\end{figure}

For Transformer-based methods, as shown in Table \ref{table:1}, the proposed LSwinSR holds a competitive performance with the SwinIR \cite{liang2021swinir} for all three scales ($ \times 2 $, $ \times 4 $ and $ \times 8 $). Taking the $ \times 2 $ scale as an example, although the SwinIR \cite{liang2021swinir} performs better on the training and validation sets, our LSwinSR delivers the best PSNR and MAE on the test set, demonstrating strong robustness and generalizability. More quantitative results for the super-resolution of the proposed LSwinSR can refer to \ref{sec:appendix}.

\subsubsection{Qualitative Comparisons} 

\begin{figure*}[htp]
\centering
\includegraphics[width=18cm]{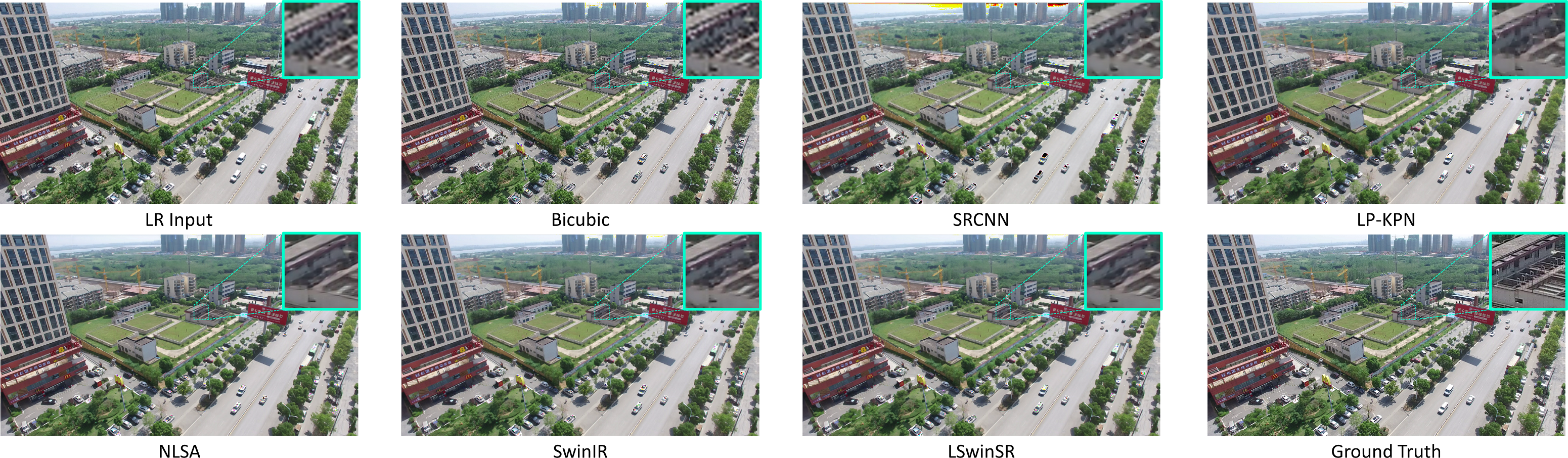}
\caption{Visual comparison of SR ($ \times 8 $) images between different methods on the test set.}
\label{fig:7}
\end{figure*}

In addition to quantitative results, the qualitative comparison of the reconstructed results with different methods is also provided from Fig. \ref{fig:2} to Fig. \ref{fig:7}, where an enlarged region is presented on the corner within each corresponding image for convenient comparison. 

Obviously, a better reconstruction can be achieved on the lower super-resolution scale. For example, in Fig. \ref{fig:2}, the e-bike riders (top) and the building contours (bottom) generated by different methods are all recognizable. However, the noises of images generated by the Bicubic, SRCNN \cite{dong2015image}, LP-KPN \cite{cai2019toward} and NLSA \cite{mei2021image} are much clearer and more than those by the SwinIR \cite{liang2021swinir} and the LSwinSR. For example, in the bottom part of Fig. \ref{fig:2}, there exist dozens of obvious reddish and blackish noisy points on the white vertical ridge of the building in images predicted by the Bicubic and CNN-based methods. In contrast, for those images generated by Transformer-based methods, the existence of such severe noisy points is much less frequent. Specifically, images predicted by SwinIR \cite{liang2021swinir} and LSwinSR contain only two and one such noisy points, respectively.

With the increase of the upsample scale, the available information for accurate reconstruction becomes more and more scarce, leading to great challenges for super-resolution algorithms. For example, in the top part of Fig. \ref{fig:4}, only SwinIR \cite{liang2021swinir} and the proposed LSwinSR provide smooth results for the text in the image when upsampling $ \times 4 $ scale. By comparison, the image generated by the Bicubic interpolation contains serious noises, while the image predicted by SRCNN \cite{dong2015image} is more blurry than others. For LP-KPN \cite{cai2019toward} and NLSA \cite{mei2021image}, the problem is the unnecessary wrinkles in the reconstructed texts. When it comes to $ \times 8 $ super-resolution, the super-resolution algorithms can only rely on very limited input information. As shown in Fig. \ref{fig:7}, the details of the building structure are nearly indistinguishable for all methods, but the Transformer-based methods still provide a clearer contour.

\subsubsection{Inference Speed} 

\begin{table}[htb]
\setlength{\abovecaptionskip}{0.cm}
\centering
\caption{The comparison of complexity and inference speed between SwinIR \cite{liang2021swinir} and LSwinSR under different window sizes. The complexity is measured by the Multiply-Accumulate Operations (MACs), while the inference speed, i.e.  Frames Per Second (FPS), is evaluated based on the $ 4 \times 3 \times 128 \times 128$ input.}
\label{table:2}
\begin{tabular}{cccc}
\hline
Window Size         & Model   & Complexity (G) $\downarrow$ & Speed $\uparrow$ \\ \hline
\multirow{2}{*}{8}  & LSwinSR & \textbf{15.3}       & \textbf{14.2}  \\
                    & SwinIR  & 16.8       & 13.5  \\ \hdashline
\multirow{2}{*}{16} & LSwinSR & \textbf{17.6}       & \textbf{12.2}  \\
                    & SwinIR  & 25.8      & 6.9   \\ \hdashline
\multirow{2}{*}{32} & LSwinSR & \textbf{26.6}       & \textbf{6.7}   \\
                    & SwinIR  & 62.0       & 2.1   \\ \hline
\end{tabular}
\end{table}

As the proposed LSwinSR is an improved and simplified version of SwinIR \cite{liang2021swinir}, the complexity and the inference speed between the two models are compared. As can be seen in Table \ref{table:2}, the complexity and the inference speed of our LSwinSR are slightly better than SwinIR \cite{liang2021swinir} when the window size is eight. The reason why the gap is so small is two-fold. First, the other operations except for the self-attention mechanism such as patch embedding, patch merging and MLP also occupy a large part of the network structure. Second, the complexity of the traditional self-attention mechanism increases quadratically with the size of the input, while the high computational requirement is not obvious for a small window size. Therefore, the complexity and inference speed gaps between the LSwinSR and SwinIR \cite{liang2021swinir} are dramatically widened with the increase of window sizes. When the window size reaches 32, the inference speed of the LSwinSR is at least three times faster than the SwinIR \cite{liang2021swinir}.

\begin{table*}[]
\setlength{\abovecaptionskip}{0.cm}
\centering
\caption{Quantitative comparison for segmentation based on super-resolution images generated by different methods on the test set. The results based on the original high-resolution images (HR) and the high-resolution image with Gaussian Blur (HR-Blur) are marked in \textcolor{purple}{light purple} and \textcolor{lightblue}{light blue} respectively, while the best result is \textbf{highlighted} and the second best is \underline{underlined}.}
\label{table:3}
\begin{tabular}{ccccccccccc}
\hline
Model    & Scale        & Clutter & Building & Road  & Tree  & Vegetation & Static Car & Moving Car & Human & mIoU  \\ \hline
Bicubic  & $ \times 2 $ & 0.592   & \underline{0.825}    & 0.761 & 0.749 & 0.577      & 0.417      & 
0.661 & 0.229 & 0.601 \\
SRCNN \cite{dong2015image}    & $ \times 2 $ & 0.602   & 0.814    & \underline{0.766} & 0.734 & 0.575      & 0.360      & 0.661      & 0.244 & 0.595 \\
LP-KPN \cite{cai2019toward} & $ \times 2 $ & 0.596   & 0.815    & 0.747 & 0.756 & 0.559      & 0.414      & 0.658      & 0.246 & 0.599 \\
NLSA \cite{mei2021image}     & $ \times 2 $ & 0.599   & 0.817    & 0.761 & 0.754 & 0.583      & 0.440      & 0.656      & 0.257 & 0.608 \\
SwinIR \cite{liang2021swinir}   & $ \times 2 $ & \textbf{0.617}   & \textbf{0.828}    & \textbf{0.771} & \underline{0.757} & \underline{0.599}      & \underline{0.441}      & \underline{0.683}      & \underline{0.265} & \underline{0.620} \\
LSwinSR  & $ \times 2 $ & \underline{0.605}   & 0.821    & 0.756 & \textbf{0.774} & \textbf{0.602}      & \textbf{0.469}      & \textbf{0.686}      & \textbf{0.273} & \textbf{0.623} \\\hdashline[0.5pt/5pt]
HR-Blur  & $ \times 2 $ & \textcolor{purple}{0.603}   & \textcolor{purple}{0.835}    & \textcolor{purple}{0.745} & \textcolor{purple}{0.705} & \textcolor{purple}{0.562}      & \textcolor{purple}{0.476}      & \textcolor{purple}{0.630}      & \textcolor{purple}{0.232} & \textcolor{purple}{0.599} \\
HR       & $ \times 2 $ & \textcolor{lightblue}{0.635}   & \textcolor{lightblue}{0.850}    & \textcolor{lightblue}{0.775} & \textcolor{lightblue}{0.777} & \textcolor{lightblue}{0.593}      & \textcolor{lightblue}{0.509}      & \textcolor{lightblue}{0.676}      & \textcolor{lightblue}{0.273} & \textcolor{blue}{0.636} \\\hdashline
Bicubic  & $ \times 4 $ & 0.594   & 0.812    & 0.755 & 0.757 & 0.580      & 0.427      & 0.648      & 0.187 & 0.595 \\
SRCNN \cite{dong2015image}    & $ \times 4 $ & 0.593   & 0.813    & 0.743 & \textbf{0.759} & \textbf{0.587}      & 0.392      & 0.643      & 0.203 & 0.592 \\
LP-KPN \cite{cai2019toward} & $ \times 4 $ & \underline{0.603}   & \underline{0.827}    & \underline{0.765} & \textbf{0.759} & 0.585      & 0.436      & 0.653      & \underline{0.206} & 0.604 \\
NLSA \cite{mei2021image}     & $ \times 4 $ & 0.585   & 0.816    & 0.755 & \textbf{0.759} & 0.582      & 0.399      & 0.645      & 0.202 & 0.593 \\
SwinIR \cite{liang2021swinir}   & $ \times 4 $ & \textbf{0.607}   & \textbf{0.829}    & \textbf{0.768} & 0.756 & 0.579      & \underline{0.483}      & \underline{0.655}      & 0.202 & \underline{0.610} \\
LSwinSR  & $ \times 4 $ & \underline{0.603}   & 0.824    & 0.761 & 0.757 & \underline{0.586}      & \textbf{0.486}      & \textbf{0.657}      & \textbf{0.213} & \textbf{0.611} \\\hdashline[0.5pt/5pt]
HR-Blur  & $ \times 4 $ & \textcolor{purple}{0.611}   & \textcolor{purple}{0.841}    & \textcolor{purple}{0.758} & \textcolor{purple}{0.698} & \textcolor{purple}{0.557}      & \textcolor{purple}{0.532}      & \textcolor{purple}{0.609}      & \textcolor{purple}{0.225} & \textcolor{purple}{0.604} \\
HR       & $ \times 4 $ & \textcolor{lightblue}{0.642}   & \textcolor{lightblue}{0.850}    & \textcolor{lightblue}{0.785} & \textcolor{lightblue}{0.782} & \textcolor{lightblue}{0.601}      & \textcolor{lightblue}{0.530}      & \textcolor{lightblue}{0.680}      & \textcolor{lightblue}{0.270} & \textcolor{lightblue}{0.643} \\\hdashline
Bicubic  & $ \times 8 $ & 0.553   & 0.774    & 0.732 & 0.726 & 0.562      & 0.322      & 0.588      & 0.007 & 0.533 \\
SRCNN \cite{dong2015image}    & $ \times 8 $ & 0.446   & 0.677    & 0.640 & 0.602 & 0.464      & 0.036      & 0.150      & 0.000 & 0.377 \\
LP-KPN \cite{cai2019toward} & $ \times 8 $ & 0.565   & \underline{0.795}    & 0.743 & 0.735 & 0.558      & \underline{0.338}      & 0.611      & 0.149 & 0.562 \\
NLSA \cite{mei2021image}     & $ \times 8 $ & 0.572   & 0.793    & 0.748 & 0.730 & 0.565      & \textbf{0.376}      & 0.611      & 0.145 & 0.567 \\
SwinIR \cite{liang2021swinir}   & $ \times 8 $ & \underline{0.573}   & 0.791    & \underline{0.751} & \underline{0.744} & \textbf{0.576}      & 0.330      & \underline{0.637}      & \textbf{0.173} & \underline{0.572} \\
LSwinSR  & $ \times 8 $ & \textbf{0.590}   & \textbf{0.803}    & \textbf{0.753} & \textbf{0.748} & \underline{0.572}      & 0.321      & \textbf{0.640}      & \underline{0.159} & \textbf{0.573} \\\hdashline[0.5pt/5pt]
HR-Blur  & $ \times 8 $ & \textcolor{purple}{0.628}   & \textcolor{purple}{0.849}    & \textcolor{purple}{0.776} & \textcolor{purple}{0.713} & \textcolor{purple}{0.567}      & \textcolor{purple}{0.535}      &\textcolor{purple}{0.675}      & \textcolor{purple}{0.229} & \textcolor{purple}{0.621} \\
HR       & $ \times 8 $ & \textcolor{lightblue}{0.654}   & \textcolor{lightblue}{0.856}    & \textcolor{lightblue}{0.800} & \textcolor{lightblue}{0.787} & \textcolor{lightblue}{0.615}      & \textcolor{lightblue}{0.546}      & \textcolor{lightblue}{0.712}      & \textcolor{lightblue}{0.292} & \textcolor{lightblue}{0.658} \\ \hline
\end{tabular}
\end{table*}

\subsection{Semantic Segmentation Performance}

As the image quality assessments such as PSNR and SSIM cannot directly reflect the reliability of the predicted images for practical applications, we further evaluate the robustness of the reconstructed images from different methods for semantic segmentation. 

\begin{figure*}[]
\centering
\includegraphics[width=18cm]{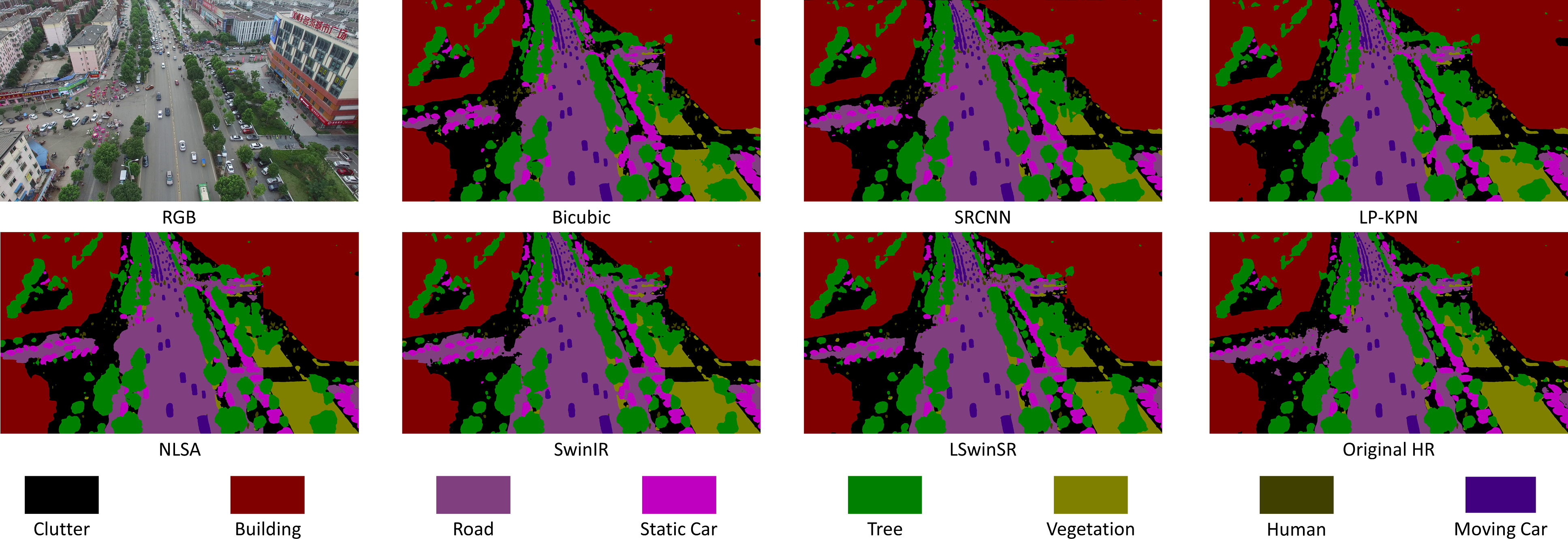}
\caption{The segmentation results on the test set based on $ \times 2 $ SR images generated by different methods.}
\label{fig:8}
\end{figure*}

\begin{figure*}[]
\centering
\includegraphics[width=18cm]{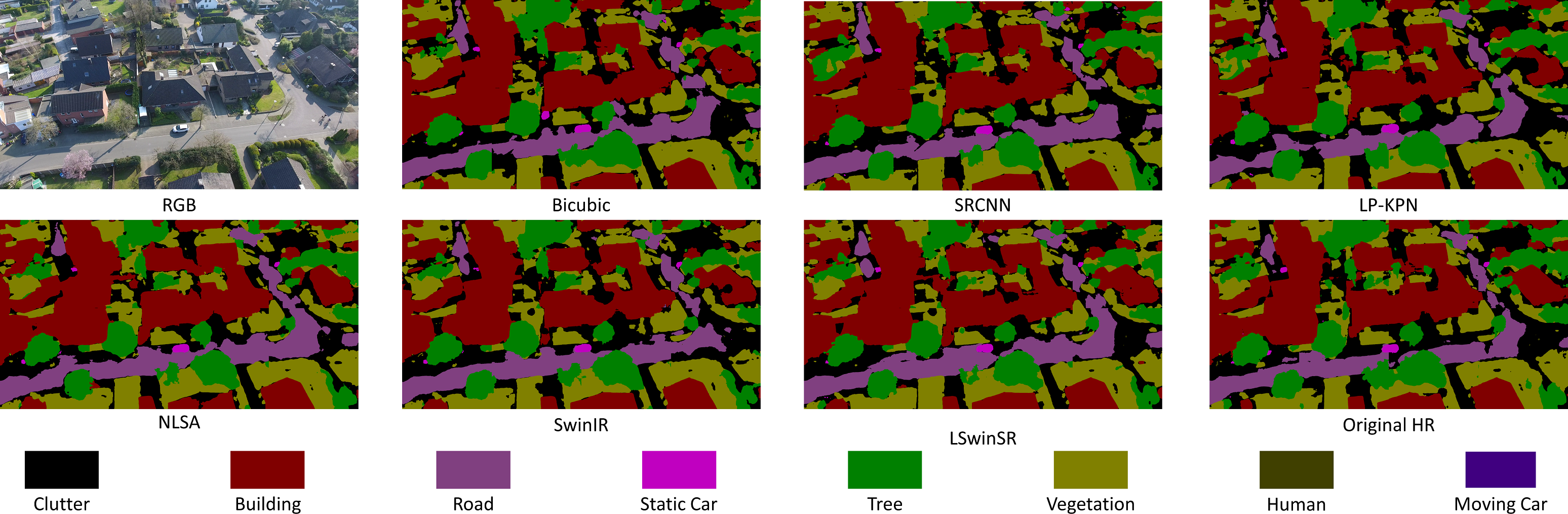}
\caption{The segmentation results on the test set based on $ \times 4 $  SR images generated by different methods.}
\label{fig:9}
\end{figure*}

\begin{figure*}[]
\centering
\includegraphics[width=18cm]{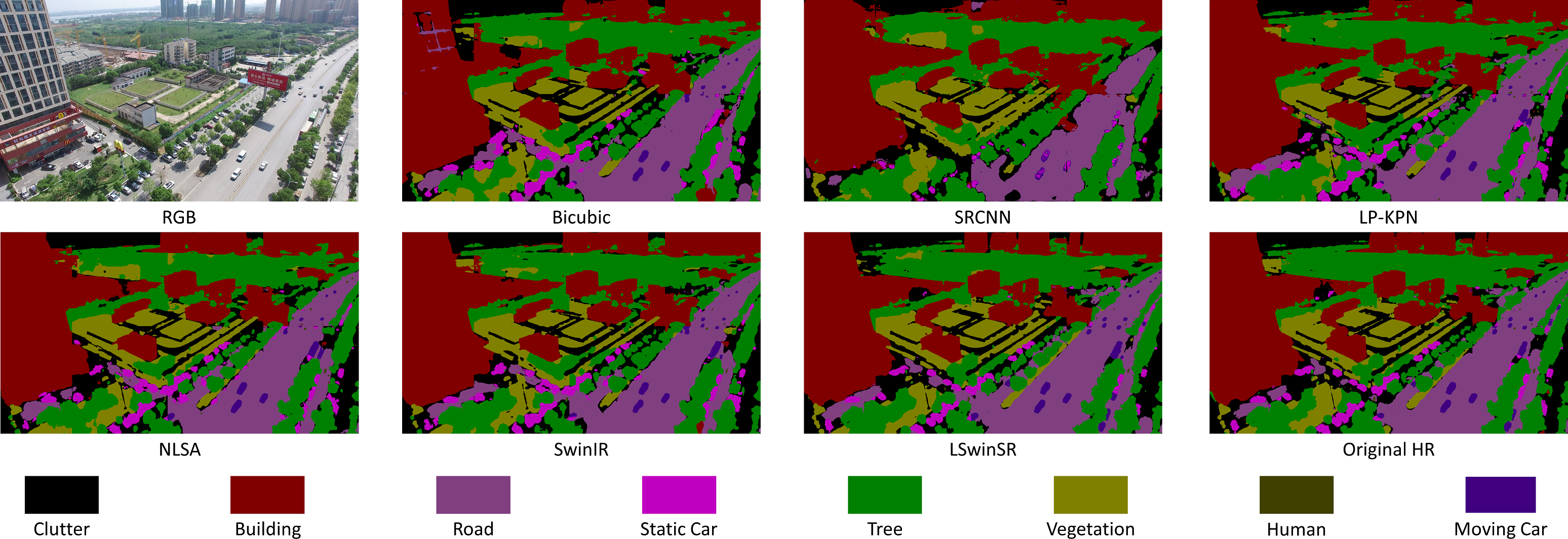}
\caption{The segmentation results on the test set based on $ \times 8 $  SR images generated by different methods.}
\label{fig:10}
\end{figure*}

For each super-resolution method, the UNetFormer \cite{wang2022unetformer} is trained and validated using the upsampled training and validation sets for segmentation, whereafter the trained UNetFormer \cite{wang2022unetformer} is then used to predict the segmentation maps of the upsampled test set, i.e. the training, validation and test sets are all upsampled by each corresponding super-resolution method. As a reference, we train the model based on the original high-resolution images and report the segmentation accuracy based on the original test set (HR) and the test set after Gaussian Blur (HR-Blur). The visual comparisons between segmentation maps based on super-resolution images generated by different algorithms are provided from Fig. \ref{fig:8} to Fig. \ref{fig:10}.

As can be seen from Table \ref{table:3}, due to the different probability distribution between the training set and test set, the accuracy of the test set with Gaussian Blur (HR-Blur) is obviously lower than the original test set (HR). Fortunately, this distributional difference can be offset by the super-resolution methods to a certain degree. For example, as shown in Table \ref{table:3}, the segmentation accuracy of the Bicubic interpolation for $ \times 2 $ scenario (mIoU: 0.601) is even better than the HR-Blur (mIoU: 0.599). We conjecture that the distributional difference gap between the training set and test set is narrowed after downsampling and upsampling operations, relieving the adverse impact caused by the Gaussian Blur. However, when it comes to the $ \times 8 $ scale, the accuracy of HR-Blur surpasses all super-resolution methods, even with a different probability distribution with the training set. The reason is that small objects such as humans and cars are hardly identifiable after downsampling to $ 1 / 8 $ resolution, while such severe information loss cannot be totally recovered by the super-resolution methods. Therefore, the adverse impact caused by information loss overshadows the distributional difference, leading to the worse segmentation accuracy of super-resolution results. 

Among five deep-learning-based methods, only the SwinIR \cite{liang2021swinir} and the proposed LSwinSR can always deliver better results than the simple Bicubic interpolation for all three scales, demonstrating the reliability and robustness of the Transformer-based super-resolution methods on different scenarios. Meanwhile, what the segmentation accuracy can show is only the superiority rather than the full potential of the Transformer-based solution. For example, as shown in the enlarged region in Fig. \ref{fig:5}, the cyclist can be clearly identified. However, only a small part of those pixels is correctly classified as Human in Fig. \ref{fig:9}, even for the original high-resolution input. The reason is that the humans in the UAVid are very small and only occupy limited pixels in the high-resolution 4K images. Without enough training data, the segmentation model naturally lacks the ability to interpret those small objects. Actually, the best accuracy on the Human object for the UAVid dataset achieved by the state-of-the-art segmentation model is only 0.33 measured by mIoU. In other words, the capability of the segmentation model limits the full potential of the super-resolution results in segmentation performance. Thus, with the development of the segmentation model especially the optimization for the identification of small objects, we believe a much better segmentation accuracy can be achieved by the super-resolution results.

It is also noteworthy that the segmentation accuracy of NLSA \cite{mei2021image} for $ \times 8 $ scale (mIoU: 0.567) is much higher than the SRCNN \cite{dong2015image} (mIoU: 0.377), although the latter holds better image quality assessments, e.g. the PSNR of SRCNN is 20.000 while the one of the NLAS is 18.453. This phenomenon can illustrate two issues. First, the higher image quality metrics are not always connected with better reliability. Second, even though the over-fitting problem can cause an extremely low image quality assessment, a tolerable segmentation result can still be guaranteed so long as the model can narrow the probability distribution gap between the training set and test set.

\section{Conclusions}
\label{sec:5}

In this work, a novel Linear Swin Transformer for Super-Resolution (LSwinSR) was proposed. The LSwinSR has addressed the high memory and computational requirements of the original Swin Transformer caused by the quadratic complexity of the embedded self-attention mechanism by introducing the kernel attention mechanism. The super-resolution experiments conducted on the large-scale UAVid dataset demonstrated that the proposed LSwinSR could provide competitive performance compared to the SwinIR \cite{liang2021swinir} but with better efficiency. Furthermore, the experiments for semantic segmentation demonstrated that the super-resolution technology could indeed enhance the segmentation accuracy, where only two Transformer-based super-resolution methods could always deliver better performance than the simple Bicubic interpolation.

In the future, we will investigate the potential solution to integrate the super-resolution models and semantic segmentation models, thereby providing more accurate and reliable segmentation results based on super-resolution technology.

\section*{Declaration of Competing Interest}
\par The authors declare that they have no known competing financial interests or personal relationships that could have appeared to influence the work reported in this paper.

\section*{CRediT authorship contribution statement}
\textbf{Rui Li}: Formal analysis, Conceptualization, Investigation, Methodology, Project administration, Software, Validation, Visualization, Writing - original draft. \textbf{Xiaowei Zhao}: Conceptualization, Formal analysis, Investigation, Methodology, Project administration, Resources, Supervision, Writing - review and editing.

\section*{Acknowledgements}
The authors acknowledge the support of the Scientific Computing Research Technology Platform (SCRTP) at the University of Warwick for providing High-Performance Computing resources.

\appendix
\renewcommand\thetable{\Alph{section}\arabic{table}}
\section{Super-resolution performance on AID}
\label{sec:appendix}
\setcounter{table}{0}    

To further demonstrate the effectiveness of the proposed LSwinSR, we conduct the experiment on the publicly available AID \cite{xia2017aid} dataset, which has been used for verifying the super-resolution performance. The AID dataset contains 10000 high-resolution images in the shape of $ 600 \times 600 $ with a 0.5m spatial resolution from 30 different types of remote sensing scenarios, such as airports, bridges and churches. For super-resolution, 7850 images are selected as the training set, 150 as the validation set and the remaining 2000 images are used as the test set. The experimental results are reported in Table \ref{table:A1}. As shown in Table \ref{table:A1}, in the large-scale and high-resolution remote sensing dataset, i.e. the AID \cite{xia2017aid}, the proposed LSwinSR can still deliver competitive accuracy compared with the existing super-resolution algorithms.

\begin{table}[htb]
\setlength{\abovecaptionskip}{0.cm}
\centering
\caption{Quantitative comparison including PSNR and SSIM (\%) with different methods on the AID dataset, where the best result is \textbf{highlighted} while the second best is \underline{underlined}.}
\label{table:A1}
\begin{tabular}{cccc}
\hline
Model   & Scale        & PSNR $\uparrow$  & SSIM $\uparrow$  \\ \hline
Bicubic & $ \times 2 $ & 32.39 & 89.06 \\
SRCNN \cite{dong2015image}   & $ \times 2 $ & 34.49 & 92.86 \\
FSRCNN \cite{dong2016accelerating}  & $ \times 2 $ & 34.73 & 93.30 \\
VDSR \cite{kim2016accurate}    & $ \times 2 $ & 35.05 & 93.46 \\
LGCNet \cite{lei2017super}  & $ \times 2 $ & 34.80 & 93.20 \\
DCM \cite{haut2019remote}     & $ \times 2 $ & 35.21 & 93.66 \\
HSENet \cite{lei2021hybrid}  & $ \times 2 $ & \underline{35.24} & \underline{93.68} \\
LSwinSR & $ \times 2 $ & \textbf{35.29} & \textbf{93.75} \\\hdashline
Bicubic & $ \times 3 $ & 29.08 & 78.63 \\
SRCNN \cite{dong2015image}   & $ \times 3 $ & 30.55 & 83.72 \\
FSRCNN \cite{dong2016accelerating}  & $ \times 3 $ & 30.98 & 84.00 \\
VDSR \cite{kim2016accurate}    & $ \times 3 $ & 31.15 & 85.22 \\
LGCNet \cite{lei2017super}  & $ \times 3 $ & 30.73 & 84.17 \\
DCM \cite{haut2019remote}     & $ \times 3 $ & 31.31 & 85.61 \\
HSENet \cite{lei2021hybrid}  & $ \times 3 $ & \underline{31.39} & \underline{85.72} \\
LSwinSR & $ \times 3 $ & \textbf{31.41} & \textbf{85.83} \\\hdashline
Bicubic & $ \times 4 $ & 27.30 & 70.36 \\
SRCNN \cite{dong2015image}   & $ \times 4 $ & 28.40 & 75.61 \\
FSRCNN \cite{dong2016accelerating}  & $ \times 4 $ & 28.77 & 77.20 \\
VDSR \cite{kim2016accurate}    & $ \times 4 $ & 28.99 & 77.53 \\
LGCNet \cite{lei2017super}  & $ \times 4 $ & 28.61 & 76.26 \\
DCM \cite{haut2019remote}     & $ \times 4 $ & 29.17 & 78.24 \\
HSENet \cite{lei2021hybrid}  & $ \times 4 $ & \underline{29.21} & \underline{78.50} \\
LSwinSR & $ \times 4 $ & \textbf{29.23} & \textbf{78.52} \\ \hline
\end{tabular}
\end{table}

\bibliography{LswinSR}

\end{document}